\documentclass[final,twocolumn,prl,aps,floats,amsfonts,amssymb,showpacs]{revtex4}
\usepackage{amsmath}
\usepackage{graphicx}

\begin{document}

\title{On the properties of steady states in turbulent axisymmetric flows}

\author{R. Monchaux}
\email{romain.monchaux@cea.fr}
\author{F. Ravelet}
\author{B. Dubrulle}
\author{A. Chiffaudel}
\author{F. Daviaud}
\affiliation{Service de Physique de l'\'Etat Condens\'e, DSM, CEA
Saclay, CNRS URA 2464,
91191 Gif-sur-Yvette, France}

\date[published in Phys. Rev. Lett. ]{}

\date{\today}


\begin{abstract}

We experimentally study the properties of mean and most probable velocity fields in a turbulent von K{\'a}rm{\'a}n flow. These fields are found to be described by two families of functions, as predicted by a recent statistical mechanics study of 3D axisymmetric flows. We show that these functions depend on the viscosity and on the forcing. Furthermore, when the Reynolds number is increased, we exhibit a tendency for Beltramization of the flow, i.e. a velocity-vorticity alignment. This result provides a first experimental evidence of nonlinearity depletion in non-homogeneous non-isotropic turbulent flow.

\end{abstract}

\maketitle
\paragraph*{Introduction}
A yet unanswered question in statistical physics is whether 
stationary out-of-equilibrium systems share any resemblance with 
classical equilibrium systems. A good paradigm to explore this 
question is offered by turbulent flows.
Incompressible flows subject to statistically stationary forcing
generally reach a steady state, in a statistical sense, 
independent of the initial conditions. Most of the current turbulence modeling is
devoted to the understanding of this state, and the role
of the velocity fluctuations or velocity smallest scales in its
construction. A distinguished feature of stationary large Reynolds
number turbulent flows is the presence of coherent structures, under
the shape of vortices in 2D \cite{basdevant81,mcwilliams90}, or vorticity
thin tubes in 3D \cite{vincent91}. These structures correspond to
regions where vorticity is locally almost aligned with velocity ---phenomenon referred to as Beltramization \cite{benzi86,farge03}. This tendency to velocity-vorticity alignment induces depletion of nonlinearities in Navier-Stokes equations. Interestingly enough, similar depletion of
nonlinearity is also observed in the inviscid counterpart of the
Navier-Stokes equation ---the so-called Euler equation--- resulting in a
slowing down of the vorticity blow-up with respect to rigorous
estimates \cite{constantin93}. A theoretical question of interest is
therefore whether turbulent flows have a natural tendency for
nonlinearity depletion, and
whether this is a characteristic of the steady states.\\
\indent The answer to the first issue is ambiguous. On one hand, a tendency for
Beltramization has indeed been observed in some numerical simulations
of stationary, nearly isotropic turbulence \cite{siggia78,pelz85}. On the other hand, similar study performed on more general flows ---such as boundary
layer \cite{rogers87}--- provided little evidence of such a property
.. A similar conclusion has been reached using
experimental data \cite{wallace92}, with slightly less reliability
owing to the difficulty to measure vorticity. As for the second
issue, it has only been partially explored in special geometries. In
2D, equilibrium states of the Euler equations have been
classified through  statistical mechanics principle by Robert and his
collaborators \cite{sommeriarobert91,chavanis03}. They indeed correspond to
Beltrami solution. In 3D axisymmetric flows ---an intermediate
situation between 2D and 3D--- a similar task has been undertaken by
Leprovost {\it et al.} \cite{leprovost05}. In the ideal case ---force free,
inviscid--- they proved the existence of an infinite number of
conserved quantities, in addition to the energy and the helicity.
They also showed the existence of an infinite number of equilibrium
states, depending on the conserved quantities. Among them, a Beltrami
state is obtained, when only the energy and the helicity are
conserved. They postulated that even in the presence of small but finite viscosity and
forcing \cite{note1}, this feature remains valid, with selection of steady
states through boundary conditions and forcing.\\
\indent The purpose of the present Letter is to perform an experimental check
of these findings. We find that mean and most probable velocity fields of axisymmetric flows
can indeed be characterized by two families of functions, as predicted
in Ref.~\cite{leprovost05}. These functions depend on the viscosity and forcing.
As the Reynolds number is increased, they evolve
towards the functions corresponding to a Beltrami state, providing a
first evidence of nonlinearity depletion in a non-homoge\-neous, non-iso\-tropic system.
\paragraph*{Theoretical background and definitions}
Consider an incompressible axisymmetric flow, with velocity 
components in a cylindrical referential ($v_r$, $v_\theta$, $v_z$). 
Due to incompressibility and axisymmetry, only two functions are 
sufficient to describe the flow, namely the angular momentum ---$\sigma(r,z)=rv_\theta$--- and either, the stream function 
$\Psi(r,z)$, such that $(v_r,v_z)=-\nabla\times \Psi e_\theta$, or 
the azimuthal vorticity $\omega_{\theta}(r,z)$. Using a variational 
method, Leprovost {\it et al.} showed that the
steady states solution of axisymmetric Euler equations, in the force-free case
obey:
\begin{equation}
\sigma = F(\Psi);\ \
\xi - \frac{F F'}{r^2}=G(\Psi); \label{eqFG}
\ \ \mbox{with}\ \ \xi = \omega_{\theta}/r,
\end{equation}

\noindent where $F$ and $G$ are arbitrary functions linked with conservation 
laws of the system ---Casimirs of $\sigma$, generalized helicity \cite{leprovost05}. For example, the steady state corresponding to 
conservation of energy and helicity is such that $F$ is linear and $G \equiv 0$, 
resulting in a linear relation between vorticity and velocity 
($\vec{v}=\lambda \vec{\omega}$). This is a Beltrami flow,  given by :

\begin{eqnarray}\label{eqBel}
v_r&=&-\frac{\pi R}{H m}J_1(m\frac{r}{R})\cos(\pi \frac{z}{H}) \nonumber \\
v_\theta&=&\frac{\lambda R}{m}J_1(m\frac{r}{R})\sin(\pi \frac{z}{H})\\
v_z&=&J_0(m\frac{r}{R})\sin(\pi \frac{z}{H})\nonumber
\end{eqnarray}

\noindent where $\lambda$ and $m$ are free parameters and $J_0$ and $J_1$ are first order
Bessel functions.\\
\indent Considering further the thermodynamics of the system, Leprovost {\it 
et al.} proved that the equilibrium states at some fixed 
coarse-grained scale are such that the most probable 
velocity flow ({\it mpvf}) is a stationary state of the Euler 
equation, {\it i.e.} satisfies Eq.(\ref{eqFG}).  Finally, they postulated 
that these force-free, inviscid results can be extended to the 
stationary states in the presence of both viscosity and forcing, 
providing the function $F$ and $G$ are selected through boundary 
conditions and forcing.

\paragraph*{Experimental setup}
In order to check the theoretical predictions, we have worked with a 
simple \lq\lq axisymmetric\rq\rq configuration: the von
K\'arm\'an flow generated by two counter-rotating impellers in a
cylindrical vessel. The cylinder radius and height are respectively
$R=100$ mm and $H=180$ mm (distance between inner faces of impellers). The impellers consist of $185$ mm diameter
disks fitted with sixteen $20$ mm high curved blades. More details about the 
experimental setup can be found in Ref.~\cite{ravelet04b}. The impellers rotation frequencies are both
set equal to $f$ to get exact counter-rotating regime. We define two forcing associated
with the concave ({\it resp.} convex) face of the blades going
forward, denoted in the sequel by direction $(-)$ ({\it resp.} $(+)$ ). 
The working fluid is either water or glycerol
at different dilution rates. The resulting accessible Reynolds numbers
($Re=2\pi f R^2 \nu^{-1}$ with $\nu$ the kinematic viscosity) vary from
$10^2$ to $3 \times 10^5$. In the exact counter-rotating regime, whatever the impellers, the flow is divided into 
two toric cells separated by an azimuthal shear layer. This setup is invariant under rotations of $\pi$ ($\cal R_{\pi}$) around any radial axis passing through the center of the cylinder. The time-averaged velocity fields we consider hereafter are $\cal R_{\pi}$-invariant and axisymmetric \cite{ravelet04b}. Velocity measurements are done with a DANTEC Laser Doppler Velocimetry (LDV) system.

\begin{figure}[h]
\begin{center}
\includegraphics[width=.25\textwidth]{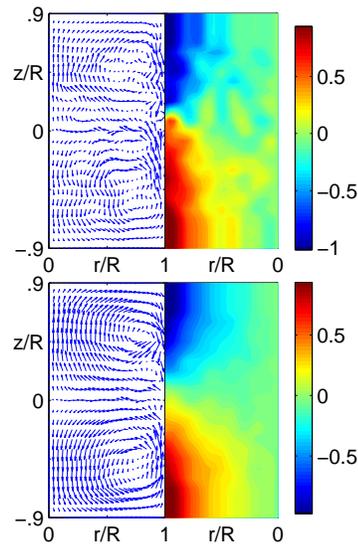}
\end{center}
\caption{Velocity profile for $Re=2.5 \times 10^5$, direction $(-)$: colored contour for azimuthal component,
arrow representation for poloidal part. Top: {\it fmpv}. Bottom:
time-averaged flow.}
\label{compareflows}
\end{figure}

\paragraph*{Data processing}
The LDV data only provide the axial and azimuthal
velocity components on a 170 points grid covering half a meridian
plane, through time series of about 200,000 randomly sampled values 
at each grid point. From this time series, it is straightforward to
get time-averaged axial and azimuthal velocities. The remaining 
radial component of the mean velocity field can then be obtained 
using the incompressibility and axisymmetry: this procedure has been 
later validated through direct measurements of the radial velocity 
with a Particles Ima\-ge
Velocimetry  (PIV) system on the same flow. We also use the time series to extract at each point histograms of the axial and azimuthal velocity, and compute the flow of most probable velocities ({\it fmpv})\cite{notefmpv}, obtained by taking, at each point, the most probable value of the velocity. The $v_r$ component is once more derived from the two other 
components  using continuity
equation. Figure \ref{compareflows} compares the two flows computed from the
same data set: they have the same overall 
structure, but differ through the size of the middle shear layer, 
which is thinner in the case of the {\it fmpv}. This is due to the 
fact that the chaotic wandering of the shear layer around its average 
position are significantly less probable as one considers positions 
further away from the equatorial plane. From a theoretical 
point of view, the two flows differ in the sense that the mean 
velocity is not a solution of the Navier-Stokes or Euler equations 
---because of the fluctuations, which induce a Reynolds stress, especially in the shear layer---, while 
the {\it mpvf} is a stationary solution of the Euler equations. Therefore, one can expect theoritical 
predictions regarding the structure of the 
stationary state to be less accurate in the case of the mean flow, as 
the Reynolds number and the fluctuations increase.

\begin{figure}[h]
\begin{center}
\includegraphics[width=.45\textwidth,clip]{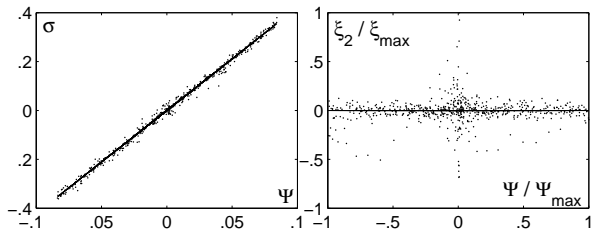}
\end{center}
\caption{$\sigma$ and $\xi_2 = \xi - \frac{F F'}{r^2}$ versus $\Psi$ for the Beltrami flow defined by equation (\ref{eqBel}) with a superimposed white noise of amplitude 60\%. We have represented the $F$ (resp. $G$) fit on the left (resp. right) figure.}
\label{Belfig}
\end{figure}

We first compute the $F$ function by looking at 
$\sigma$ as a function of $\Psi$, and then, we use this estimate to 
evaluate $G$, by looking at $\xi_2 =\xi - \frac{F F'}{r^2}$ as a function of 
$\Psi$. This procedure is likely to induce a lot of noise in the 
estimate of $G$. To check its robustness, we have tested it on a 
Beltrami  flow (Eq.~\ref{eqBel}) with a superimposed level of noise 
comparable to the level of fluctuations in the flow (see Figure 
\ref{Belfig}). Even in the presence of noise the fits give the correct 
shapes: $F=\lambda \Psi$ and $G\equiv 0$.

\begin{figure}[h]
\begin{center}
\includegraphics[width=.5\textwidth,clip]{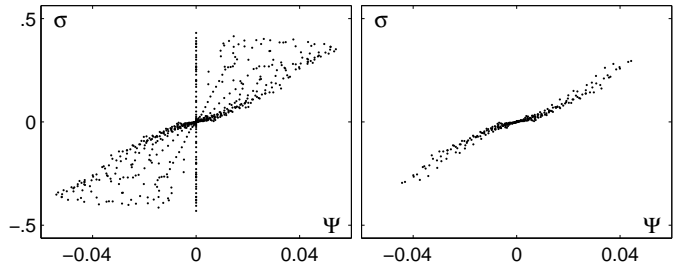}
\end{center}
\caption{$\sigma\ vs\ \Psi$ for experimental time-averaged flow in direction $(+)$ at
$Re=2100$. Left: for the whole flow. Right: for $r\leq 0.81, \
-0.56\leq z\leq 0.56$, corresponding to $50\%$ of the flow volume. The
remaining points are clearly gathering along a cubic-shaped function $F$.}
\label{naiveFa}
\end{figure}

After this test, we apply this procedure on the real data, with a 
further correction, motivated by the following remark. Figure 
\ref{naiveFa} displays
two typical plot of $F$ : one over the whole apparatus, and one over 
a $50\%$ portion of the flow obtained after removing regions close to 
the impellers and along the vessel. These parts correspond to 
locations where the viscosity and the forcing (neglected in Ref.~\cite{leprovost05}) are principally at work, and larger deviations 
from theoretical predictions can be expected. One sees that while 
data on the whole vessel display significant scatter, preventing the 
outcome of a well-defined $F$, the restricted data gather onto a 
cubic-shaped function fitted by a two parameters cubic: $F(\Psi)=p_1 
\Psi + p_3 \Psi ^3$. This fit is then used to obtain $G$. In the sequel, we obtain $F$ and $G$ 
through fits over this $50\%$ portion of the flow away from the 
boundaries.

\begin{figure}[h]
\begin{center}
\includegraphics[width=.5\textwidth]{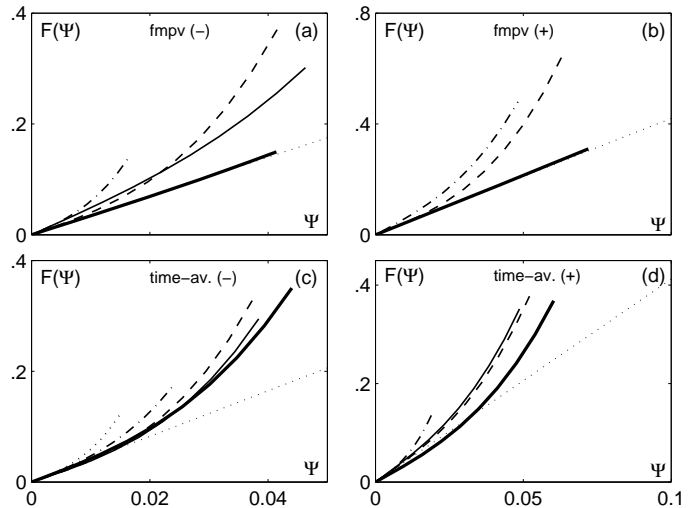}
\end{center}
\caption{F fit. $(a)$ and $(b)$: {\it fmpv} direction
$(-)$ and $(+)$ respectively, $(c)$ and $(d)$: time-averaged field  direction
$(-)$ and $(+)$ respectively.
Legend for all figures: 
$Re=100 \ (thin \ \cdot \cdot \cdot)$, 
$Re=150 \ (thin \ \cdot - \cdot)$, 
$Re=2100 \ (thin \ --)$, 
$Re=9100 \ (thin \ $---$)$, 
$Re=2.5 \times 10^5 \ (thick \ $---$)$. 
The thin dotted straight line is a Beltrami. As $F$ is 
odd, we display it for positive values of $\Psi$ only. Note that the fit at $Re=9100$ for {\it fmpv} case in direction $(+)$ could not be obtained from the data due to large scattering.}
\label{Ffit}
\end{figure}

\paragraph*{Results}
We present results computed for both time-averaged flow and {\it 
fmpv} for the two directions of rotation, at different Reynolds 
numbers. Figure \ref{Ffit} presents the $F$ fits obtained in each 
case. We have included on each graph a straight line tangent to experimental curves for $\Psi=0$ as a reference to a Beltrami flow. The experimental data are much more 
scattered in the {\it fmpv} case, debasing the fit accuracy compared to the mean flow ---$R^2 \simeq 0.94$ instead of $0.99$.

\begin{figure}[h]
\begin{center}
\includegraphics[width=.5\textwidth]{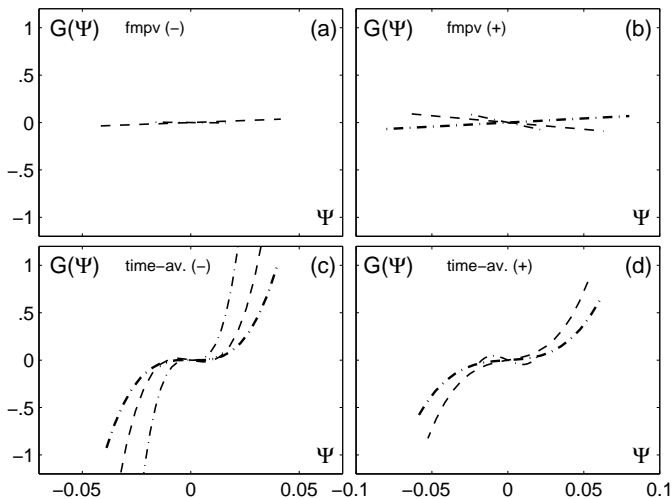}
\end{center}
\caption{G fit. $(a)$ and $(b)$: {\it fmpv} direction
$(-)$ and $(+)$ respectively, c and d: time-averaged field  direction
$(-)$ and $(+)$ respectively. Same legend as Fig.~\ref{Ffit}.}
\label{Gfit}
\end{figure}

In figure \ref{Gfit} we present the $G$ fits obtained in the
same conditions. The best fit of the {\it fmpv} $G$ functions is linear
whereas cubic-shaped functions are required for time averaged $G$. 
The plots of $G$ consist of wide noisy bands surrounding the fits in 
both cases. The noise for $G$ is larger than for $F$, since it is a 
result of a two-step procedure including a spatial derivation.

Considering the variation with Reynolds number of $F$, both the {\it fmpv} case and the time-averaged case behave similarly. As Reynolds number increases, the $F$ cubic curves collapse on the Beltrami line.

At $Re=2.5\times 10^5$, it is not even possible to distinguish the $F$ fit from a straight line and this for the two directions of rotation. In the case of $G$, there is a difference between the two cases. For the {\it fmpv}, the best fit is linear, with a slope smaller than the noise level, this is consistent with $G \equiv 0$. For the time-averaged case, the shape of $G$ is very different. All the fits  present a {\it plateau} around zero. The greater the Reynolds number, the wider this {\it plateau}, so the wider the range of $\Psi$ ---or equivalently the volume of considered flow--- where $G$ is very close to zero, {\it i.e.}, very close to the Beltrami $G$ function.

The difference in the shape of $G$ can be explained as follows : 
according to equation (\ref{eqFG}), $G(\Psi)$ is the difference of 
two terms. One linked to $F$ is necessarily a cubic, while the 
second, connected to the azimuthal vorticity is almost a noise in 
this area of the flow. In the {\it fmpv} case, the magnitude of the 
cubic term is  small compared to the vorticity, and we only see the 
noise coming from the vorticity. In the time-averaged case, the cubic 
is much larger, so that it determines the behavior of $G$.

\begin{figure}[h]
\begin{center}
\includegraphics[width=.5\textwidth]{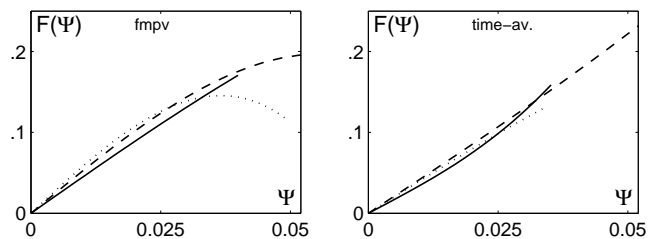}
\end{center}
\caption{F fit for $100$mm diameter impellers fitted with straight blades. Left: {\it fmpv}, Right: time-averaged. 
$Re=1.9 \times 10^5$ $(\cdot \cdot \cdot)$, 
$Re=2.6 \times 10^5$ $(- -)$, 
$Re=5 \times 10^5$ $($---$)$.}
\label{TM90}
\end{figure}

A second observation is the dependence of $F$ and $G$ on the forcing. 
Comparison of Fig.~\ref{Ffit}-(a,c) and \ref{Ffit}-(b,d), obtained for two different 
forcing, shows that the slope of $F$ slightly depends on the forcing 
($p_1=5$ for $(-)$ direction, $p_1=4$ for $(+)$ direction). For any forcing, $G$ remains close to zero at the measurement scale. To check this dependence further, we have conducted an 
additional experiment at high Reynolds number, with much smaller impellers. As shown in 
Figure \ref{TM90}, we obtained a different shape for $F$, with a 
different inflection.

\paragraph*{Discussion and perspectives}

An inviscid force-free the\-ory developed by Leprovost {\it et al.} 
predicted a characterization of the steady state in axisymmetric 
turbulent flows through two functions $F$ and $G$. In our experiments, 
we have confirmed this, and measured these functions 
for different forcing, and different viscosities, using two different 
fields as diagnostic: the {\it fmpv} as predicted and more surprisingly the time-averaged field, even in a region where Reynolds stresses are not negligible. This could be due to the quasi Gaussian distributions of velocities.

First, we have observed that the functions are well-defined only in 
the portion of the flow remote from the boundaries, where forcing and 
dissipation take place. The steady states of the von K{\'a}rm{\'a}n flow can be described by a suitable 
statistical analysis of equivalent equilibrium states (e.g. assuming 
zero viscosity and no forcing, as in Ref.~\cite{leprovost05}). 
Nevertheless, we have seen that forcing and dissipation do influence 
the steady state regime through the selection of the characteristic 
functions. Specifically, we have observed that the forcing selects 
the specific shape of the functions (linear, cubic, ...), while 
dissipation acts as a sort of self-similar zoom and select the 
portion of the curve actually explored by the flow. This suggests 
that stationary out-of-equilibrium systems like turbulent flows are 
universal in a weaker sense than in ordinary equilibrium systems : 
they can be described in a universal manner through general functions 
(like $F$ and $G$) determining the \lq\lq equation of state\rq\rq. However, 
these functions are non-universal since they depend on the fine 
details of the system (dissipation and forcing).

Finally, we have shown that the evolution for increasing Reynolds 
number is towards a Beltrami state, with depletion of 
nonlinearities. This evolution is more obvious when considering the 
{\it fmpv}, as expected from the thermodynamics analysis developed by 
Leprovost {\it et al.}. To our knowledge, this is the first 
experimental evidence of nonlinearity depletion in a non-homogeneous, 
non-isotropic turbulence, where it is very challenging to measure 
simultaneously all components of velocity and vorticity. Our use of 
the functions characterizing the steady states enables us to lower 
the experimental constraints for such a check.

\end{document}